\documentclass[aps,twocolumn,amsmath,amssymb,groupedaddress,pra]{revtex4}
\pdfoutput=1

\usepackage{graphicx}

\newcommand{\ket}[1]{\mbox{$\vert #1 \rangle$}}
\newcommand{\bra}[1]{\mbox{$\langle #1 \vert$}}
\newcommand{\up}{\mbox{$\vert 0 \rangle$}}
\newcommand{\down}{\mbox{$\vert 1 \rangle$}}
\begin{document}

\title{Quantum Walk in Position Space with Single Optically Trapped Atoms}

\author{Micha\l\ Karski}
\email[\emph{Electronic address: }]{karski@uni-bonn.de}
\author{Leonid F\"{o}rster}
\author{Jai-Min Choi}
\author{Andreas Steffen}

\author{Wolfgang Alt}
\author{Dieter Meschede}
\author{Artur Widera}
\email[\emph{Electronic address: }]{widera@uni-bonn.de}

\affiliation{
  Institut f\"{u}r Angewandte Physik der Universit\"{a}t Bonn,
  Wegelerstra\ss e 8,
  53115 Bonn,
  Germany}

\begin{abstract}
The quantum walk is the quantum analogue of the well-known random walk, which forms the basis for models and applications in many realms of science.
Its properties are markedly different from the classical counterpart and might lead to extensive applications in quantum information science. In our experiment, we implemented a quantum walk on the line with single neutral atoms by deterministically delocalizing them over the sites of a one-dimensional spin-dependent optical lattice. 
With the use of site-resolved fluorescence imaging, the final wave function is characterized by local quantum state tomography, and its spatial coherence is demonstrated. Our system allows the observation of the quantum-to-classical transition and paves the way for applications, such as quantum cellular automata.\end{abstract}

\maketitle
Interference phenomena with microscopic particles are a direct consequence of their quantum-mechanical wave nature  \cite{mllenstedt_beobachtungen_1956,carnal_youngs_1991,chapman_photon_1995,weitz_multiple_1996,hackermuller_decoherence_2004}. The prospect to fully control quantum properties of atomic systems has stimulated ideas to engineer quantum states that would be useful for applications in quantum information processing, for example, and also would elucidate fundamental questions, such as the quantum-to-classical-transition \cite{schlosshauer_decoherence_2007}. A prominent example of state engineering by controlled multipath interference is the quantum walk of a particle \cite{kempe_quantum_2003}. Its classical counterpart, the random walk, is relevant in many aspects of our life providing insight into diverse fields: It forms the basis for algorithms \cite{barber_random_1970}, describes diffusion processes in physics or biology \cite{barber_random_1970, berg_random_1993}, such as Brownian motion, or has been used as a model for stock market prices \cite{Fama1965}. Similarly, the quantum walk is expected to have implications for various fields, for instance, as a primitive for universal quantum computing \cite{childs_2009}, systematic quantum algorithm engineering \cite{dr_quantum_2002} or for deepening our understanding of the efficient energy transfer in biomolecules for photosynthesis \cite{sension_biophysics:_2007}.

Quantum walks have been proposed to be observable in several physical systems \cite{travaglione_implementingquantum_2002,dr_quantum_2002,knight_quantum_2003}. Special realizations have been reported in either the populations of nuclear magnetic resonance samples \cite{du_experimental_2003,ryan_experimental_2005}; or in optical systems, in either frequency space of a linear optical resonator \cite{bouwmeester_optical_1999}, with beam splitters \cite{do_experimental_2005}, or in the continuous tunneling of light fields through waveguide lattices \cite{perets_realization_2008}. Recently, a three-step quantum walk in the phase space of trapped ions has been observed \cite{schmitz_quantum_2009}. However, the coherent walk of an individual quantum particle with controllable internal states as originally proposed by Feynman \cite{feynman_1965} has so far not been observed.
We present the experimental realization of such a single quantum particle walking in a one-dimensional (1D) lattice in position space. This basic example of a walk provides all of the relevant features necessary to understand the fundamental properties and differences of the quantum and classical regimes. For example, the atomic wave function resulting from a quantum walk exhibits delocalized coherence which reflects the underlying quantum interference. 
Simultaneous detection of internal state and the atomic position in the lattice by an optical microscope allows for local quantum state tomography of the wave function. 
This is an important requirement to realize applications in quantum information science such as the quantum cellular automaton \cite{Raussendorf_quantum_2005,shepherd_universally_2006,vollbrecht_reversible_2006}.

In the classical random walk on a line, a coin is tossed in each time step. Depending on the outcome (heads or tails) a walker takes one step to the left or to the right. After $N$ time steps, the probability of finding the walker at a certain site on the line follows a binomial distribution with a width increasing proportional to $\sqrt{N}$. 

\begin{figure*}
	\centering
		\includegraphics[scale=0.5]{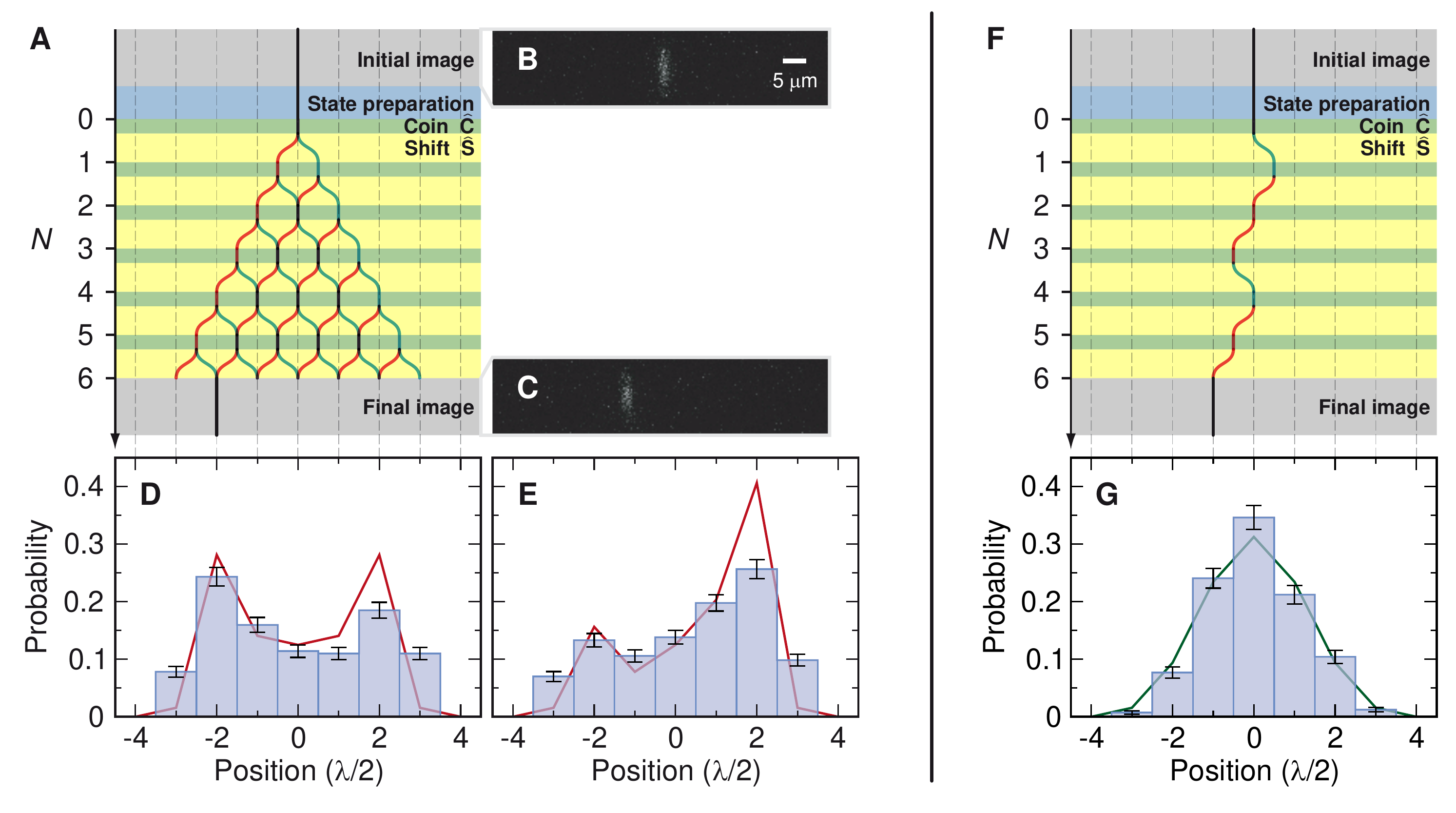}
	\caption{((A) Schematic experimental sequence for the quantum walk showing the paths for the internal states \up\ (green) and \down\ (red).  The walking distance is extracted from the initial (B) and final (C) fluorescence image. The results of several hundreds of identical realizations form the probability distribution, which is symmetric for the initial state $(\up + i \down)/\sqrt{2}$ (D) and anti-symmetric for the initial state \down\ (E).
	The analogous random walk sequence (F) yields a binomial probability distribution (G). The displayed path is one of many random paths that the atom can take. Measured data is shown as a histogram, the theoretical expectation for the ideal case as a solid line; error bars indicate the statistical $\pm 1 \sigma$ uncertainty.}
	\label{fig:Fig1}
\end{figure*}
In the quantum case, the walker is brought in a coherent superposition of going to the right or left. This can be realized by adding internal states to the walker, providing an additional degree of freedom, which can be used to control the system.
We consider a two-level particle with internal states \up\ and \down. 
In every step of the walk, the coin operator brings each internal state into a coherent superposition of the two states. The essence of the general quantum walk is to entangle this internal state with the position of the corresponding wave packet by a state-dependent transport. This can be realized by shifting both internal states into opposite directions, which coherently delocalizes the particle over two lattice sites. Repetition of the unitary coin--shift operation sequence results in the so-called quantum walk. After two steps of the quantum walk two parts of the wave function are re-combined at a common lattice site. Being in different internal states they cannot interfere. The next coin operator, however, mixes the internal states in a deterministic way, which gives rise to quantum interference of the two overlapping wave packets. Further steps result in a multipath interference (Fig.~\ref{fig:Fig1} A) which then alters the properties of the quantum walk compared to the classical random walk. 
In particular, the width of the probability distribution to find the walker at a certain position scales proportional to $N$ for the quantum walk, as in a ballistic transport, in contrast to the diffusive $\sqrt{N}$ scaling of the random walk. The influence of internal states on the quantum walk provides another distinguishing feature: Whereas the probability distribution of the random walk is fully determined by the balance of the coin, the quantum walk distribution strongly depends on the initial internal state of the walker and can be either symmetric or strongly asymmetric for one and the same coin operator (see Fig.~\ref{fig:Fig1}). Furthermore, as the quantum walk is fully deterministic and unitary, the multipath interference can be reversed by inverting coin and shift operations.
\begin{figure*}
	\centering
		\includegraphics [scale=0.4]{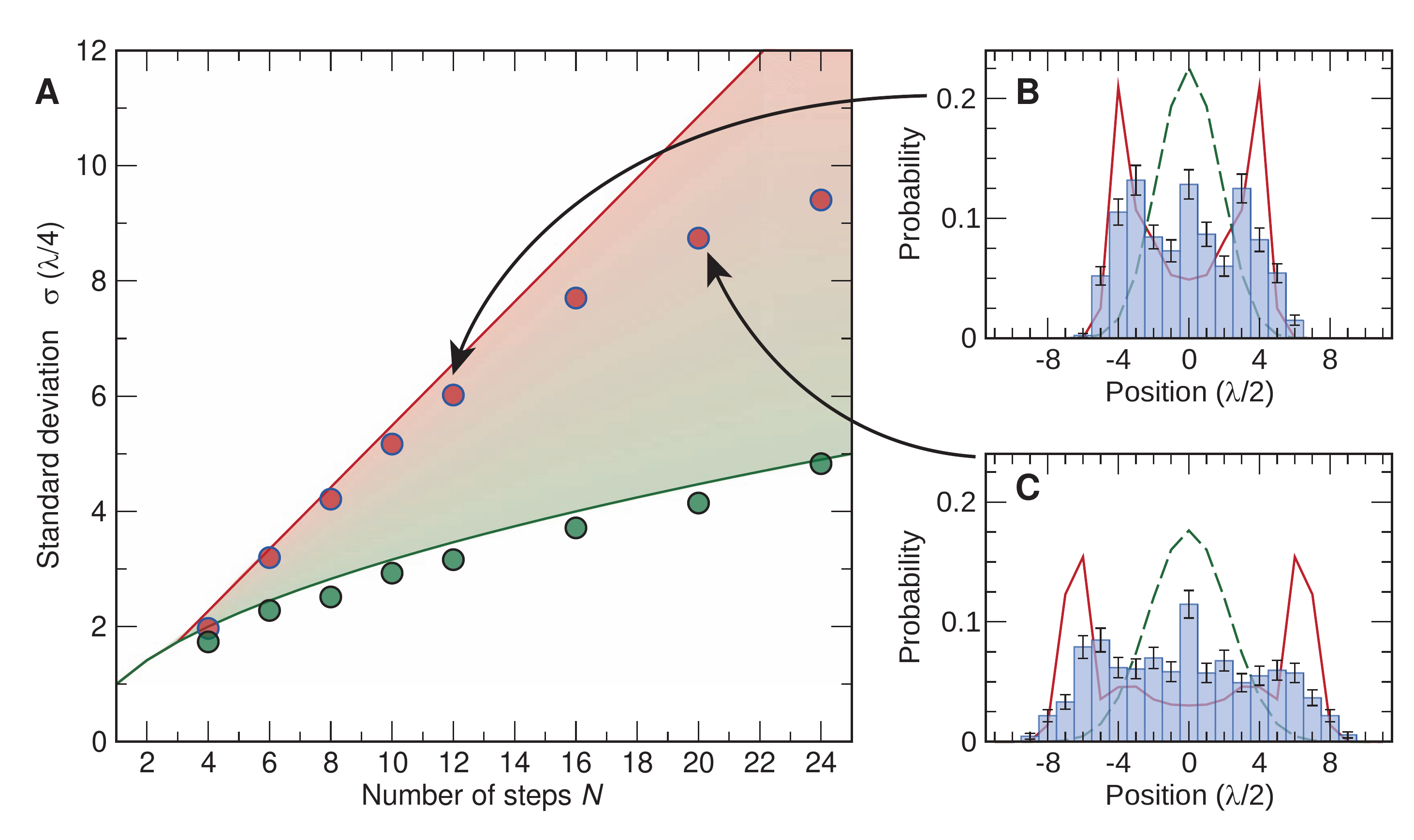}
	\caption{(A) Scaling of the standard deviation of the measured spatial probability distributions for quantum walk (red) and random walk (green). The solid lines indicate the expectations for the ideal cases. Error bars are smaller than the size of symbols. The measured quantum walks follow the ideal linear behavior until, because of decoherence, they gradually turn into a random walk.
	 The probability distributions for $N=12$ (B) and $N=20$ (C) show a gradual change from the quantum to a classical shape. The theoretical prediction is shown as a solid line for the pure quantum walk and as a dashed line for the random walk.}
	\label{fig:Fig2}
\end{figure*}

We realize a quantum walk with single laser-cooled cesium (Cs) atoms, trapped in the potential wells of a 1D optical lattice\cite{dr_quantum_2002} with site separation of $\lambda/2 = 433\,$nm (here $\lambda$ is the wavelength of the lattice laser light). The atoms are thermal with a mean energy of $k_B \times 10\,\mu$K, whereas the optical potential depth is $k_{\mathrm{B}} \times 80\,\mu$K (here, $k_\mathrm{B}$ is the Boltzman constant). They are distributed among the axial vibrational states with a mean occupation number of $\bar{n}_\mathrm{ax} = 1.2$. Initially, the atoms are prepared in the $\up \equiv \ket{F=4,m_F=4}$ hyperfine state by optical pumping, where $F$ is the total angular momentum and $m_F$ its projection onto the quantization axis along the dipole trap axis. Resonant microwave radiation around $9.2\,$GHz coherently couples this state to the $\down \equiv \ket{F=3, m_F=3}$ state.
A $\pi/2$-pulse of $4\,\mu$s initializes the system in the superposition $(\up + i \down)/\sqrt{2}$. 
Coin operators are realized in form of Hadamard-type gates
\begin{equation} 
 \hat{C}: \left\{ \begin{array}{c}\up \rightarrow \up - \down)/\sqrt{2}  \\ \down \rightarrow (\up + \down)/\sqrt{2} \end{array} \right. . 
\end{equation}
The state-dependent shift operation is performed by continuous control of the trap polarization, moving the spin state \up\ (\down) adiabatically to the right (left) along the lattice axis within $19\,\mu$s (see Supplementary Material). After $N$ steps of coin operation and state-dependent shift, the final atom distribution is probed by fluorescence imaging. From these images the exact lattice site of the atom after the walk is extracted \cite{karski_nearest-neighbor_2009} and compared to the initial position of the atom. Spin echo operations are combined with each coin operation (see Supplementary Material), leading to a coherence time of 0.8\,ms.

The final probability distribution $P_N(\xi)$ to find an atom at position $\xi$ after $N$ steps (see Fig.~\ref{fig:Fig1}) is obtained from the distance each atom has walked by taking the ensemble average over several hundreds of identical realizations of the sequence. Ideally, one expects a double-peak distribution with large amplitude close to the edges of the distribution \cite{kempe_quantum_2003}. The relative heights of the left and right peaks -- and therefore the symmetry -- depend on the choice of the initial state. Decoherence gradually suppresses the pronounced peaks \cite{dr_quantum_2002,kendon_decoherence_2003}. We compare the measured distributions for the symmetric and asymmetric quantum walks of $N=6$ steps (Fig.~\ref{fig:Fig1} D,E) with the theoretical expectations for the ideal case and find good agreement. 

In contrast, a random walk distribution can be recovered by introducing decoherence after each step of the walk. Omitting the spin-echo from the coin operation and additionally waiting 400\,$\mu$s between coin and subsequent shift operation destroys the phase relation between subsequent steps of the walk. The resulting probability distribution is described by a binomial distribution (Fig.~\ref{fig:Fig1} G), as expected for a purely classical random walk.

\begin{figure*}
	\centering
		\includegraphics [scale=0.7]{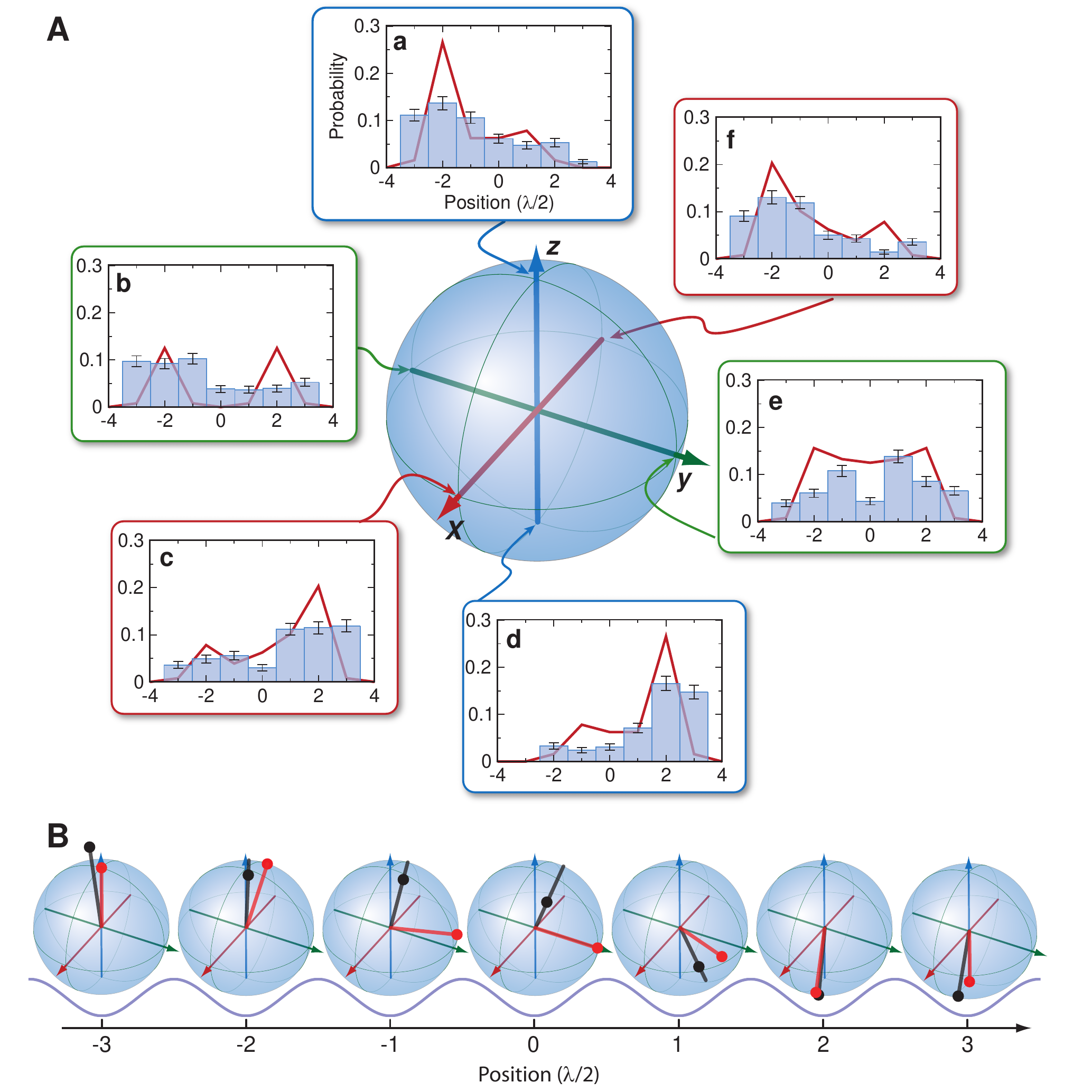}
	\caption{(A) Local quantum state tomography of the atomic wave function after a six-step quantum walk. The distributions belong to the eigenstates of the Pauli spin operators $\hat{\sigma}_i$ ($i= x,y,z$): (a) $\up$ ($+z$-axis), (b) $(\up - i \down)/\sqrt{2}$ ($-y$-axis), (c) $(\up + \down)/\sqrt{2}$ ($+x$-axis), (d) $\down$ ($-z$-axis), (e) $(\up + i \down)/\sqrt{2}$ ($+y$-axis), and (f) $(\up - \down)/\sqrt{2}$ ($-x$-axis ). (B) Reconstructed Bloch vectors at each position in the lattice. The tips of the reconstructed and ideally expected Bloch vectors are shown as black and red dots, respectively. 
The lines for Bloch vectors extend to the surface of the Bloch sphere to guide the eye; deviations from the surface illustrate the effect of decoherence and measurement errors. }
	\label{fig:Fig3}
\end{figure*}
The scaling of the width of the quantum and the random walk distribution with the number of steps is one of the most prominent distinguishing features. We have investigated this scaling behavior for both walks for up to $N=24$ steps (Fig.~\ref{fig:Fig2}). For the quantum walk, the width follows closely the expected linear behavior for up to 10 steps. The subsequent deviation is due to decoherence (for details see Supplementary Material), which asymptotically turns the quantum walk into a classical random walk. In contrast, for the random walk the typical square-root scaling is recovered.
\begin{figure*}
	\centering
		\includegraphics [scale=0.5]{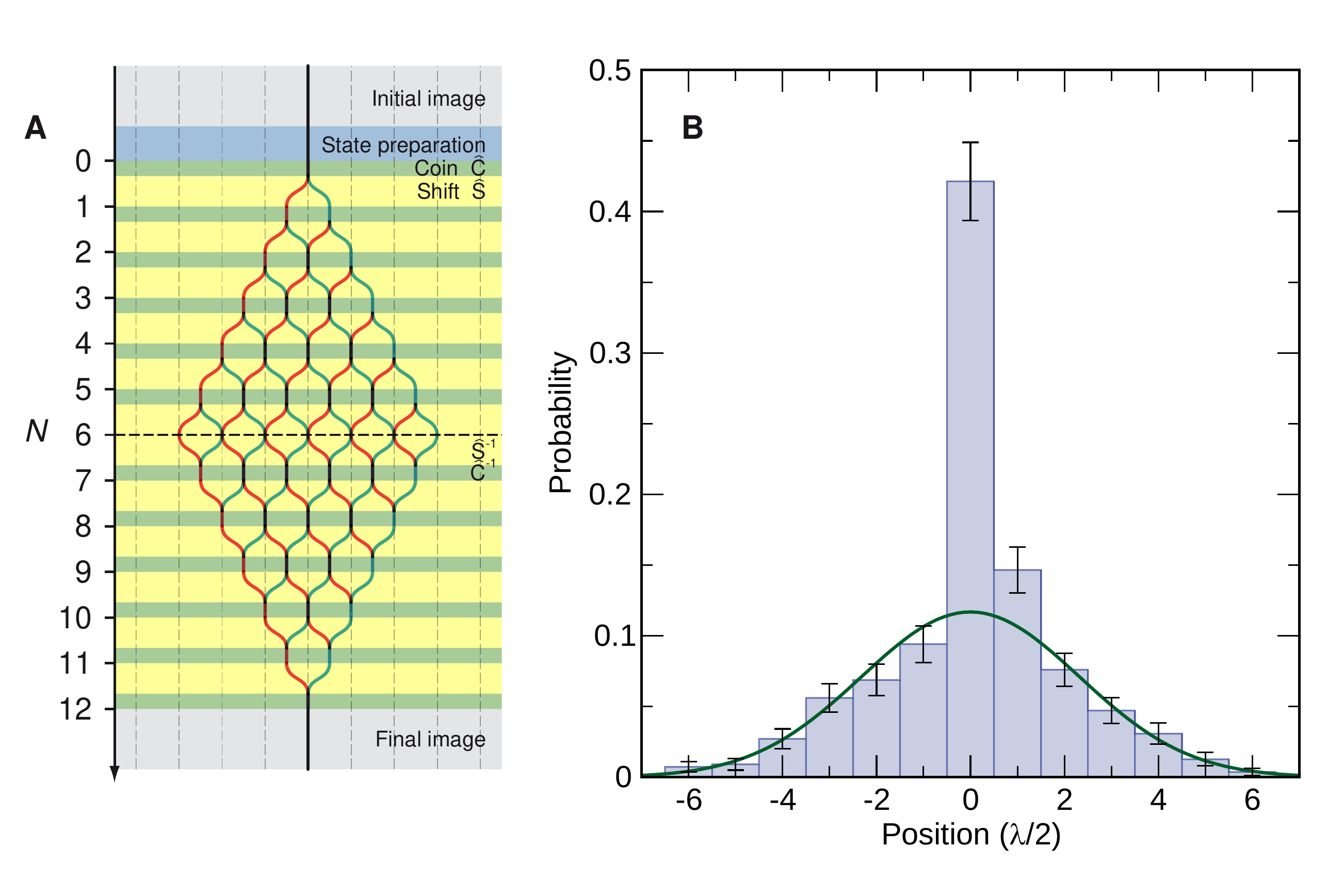}
	\caption{(A) Time-reversal sequence for re-focussing the delocalized state of a six-step quantum walk. After six steps, the total application of the coin and shift operator is reversed, where ($\hat{S}\, \hat{C})^{-1} = \hat{C}^{-1} \, \hat{S}^{-1}$. (B) The resulting probability distribution shows a pronounced peak at the center, where ideally the amplitude should be fully refocussed to. We observe a refocussed amplitude of 30\%, surrounded by a Gaussian background (fitted curve).}
	\label{fig:Fig4}
\end{figure*}
To get a more detailed characterization of the wave function prepared by a six-step quantum walk sequence, we extract information on the internal state populations and relative phase by local quantum state tomography. This is based on site-resolved, state-selective detection combined with single-particle operations \cite{rosenfeld_remote_2007} (see Supplementary Material), providing a population distribution for each eigenstate of the Pauli spin operators $\hat{\sigma}_i$ ($i=x,y,z$), see Fig.~\ref{fig:Fig3}. Essentially, at each lattice site, the internal quantum state is represented by a vector on the Bloch sphere, which we reconstruct from the result of the tomography. These Bloch vectors fit well to the theoretical prediction at the edges of the distribution, but they show increasing deviations in a region close to the initial site of the walk. At these lattice sites, matter wave interference occurs at almost every step during the sequence, which makes these lattice sites more sensitive to decoherence compared with sites further apart.

The local tomography, however, does not yield information about the off-diagonal elements of the position space density matrix, which essentially contain information about the phase relation between the wave function at different lattice sites rather than at each site. To demonstrate the spatial coherence of the state over all populated lattice sites, we invert the coin operation 
\begin{equation}
C^{-1}: \left \{ \begin{array}{c} \up \rightarrow \up + \down)/\sqrt{2} \\ \down \rightarrow (\up - \down)/\sqrt{2} \end{array} \right. 
\end{equation}
as well as the shift operation, and continue the walk for six additional steps (see Fig.~\ref{fig:Fig4}). Ideally, the inversion acts as an effective time-reversal and refocusses the multipath interference pattern of the wave function back to the initial lattice site. We find partial refocussing of 30\% of the atomic population to the expected lattice site reflecting the fraction of atoms which have maintained coherence throughout the sequence.

We have studied the quantum walk of single neutral atoms in an optical lattice and characterized the quantum state of the delocalized atom. We have found good agreement with the ideal case of a quantum walk for up to ten steps. Inversion of the walk causes the delocalized wave function to refocus to the initial lattice site.
Although the atoms in our experiments are thermally distributed among several vibrational states, we obtain large coherence over a macroscopic distance. In the ideal case, motional state and internal states factorize, so that the coherence created in one degree of freedom is not affected by the other. We have found that, as soon as internal and external degrees of freedom are coupled by diabatic transport leading to vibrational excitations, for instance, the matter wave interference is quickly suppressed. 

It will be interesting to investigate the behavior of quantum walks for different conditions when coin operations depend on position or time. In particular, monitoring the decay of coherence under the influence of different noise sources will further elucidate the transition from the quantum to the classical regime. Performing the quantum walk with more than one atom and enabling coherent interactions between the atoms\cite{mandel_controlled_2003} will realize first operational quantum cellular automata that can be probed by full quantum state tomography, opening another experimental route towards quantum information science.

We thank D. D\"oring, F. Grenz, and A. H\"arter for help in the construction of the apparatus and A. Rauschenbeutel for valuable discussions. We acknowledge financial support from the DFG (research unit 635) and the European Comission (Integrated Project on Scalable Quantum Computing with Light and Atoms). M.~Karski acknowledges support from the Studienstiftung des deutschen Volkes, J.-M.~Choi received partial support from the Korea Research Foundation Grant funded by the Korean Government (Ministry Of Education and Human Resources Development).

\subsection*{Supplementary Material}
\paragraph*{Trapping single atoms}

A small number of Caesium atoms is transferred from a magneto-optical trap into the periodic dipole potential of a standing-wave laser field –-- a 1D optical lattice. The optical lattice potential is created by two counter propagating Gaussian laser beams at a wavelength of $\lambda=865.9\,$nm. A waist of $w_0=20\,\mu$m and a typical power of $P=20\,$mW provide a trap depth of $k_{\mathrm{B}}\times 80\,\mu$K. The corresponding axial and radial trapping frequencies of the atoms are $\omega_{\mathrm{ax}} = 2\pi\times 120\,$kHz and $\omega_{\mathrm{rad}}= 2\pi\times 1.5\,$kHz. In the lattice, the atoms are Doppler cooled by a red detuned three-dimensional optical molasses at $852\,$nm which is also used to illuminate the atoms for the fluorescence imaging. The atomic temperature is $k_{\mathrm{B}}\times 10\,\mu$K, corresponding to mean vibrational quanta of $\bar{n}_{\mathrm{ax}} = 1.2$ axially and $\bar{n}_\mathrm{rad} = 200$ radially.

\paragraph*{Shift operation}
For a lin-$\theta$-lin polarization configuration at a wavelength of $\lambda=865.9\,$nm, the two internal states \up\ and \down\ couple to orthogonal circularly polarized light fields. If the angle $\theta$ between the linear polarizations is rotated, the resulting standing wave light field can be decomposed into a right- and a left-hand circularly polarized standing wave with a relative displacement $\Delta x = (\theta \,\lambda)/(2 \pi)$ \cite{brennen_quantum_1999,jaksch_entanglement_1999}. The two internal states \up\ and \down\ experience the respective potentials 
\begin{eqnarray}
U_{\scriptsize \up} &=& U_\mathrm{+} \\
U_{\scriptsize \down} &=& \frac{7}{8} \, U_\mathrm{-} + \frac{1}{8}\, U_\mathrm{+},
\end{eqnarray}
where 
\begin{equation}
U_{\pm} = V_0\,\cos^2\left( k\, x \pm \Delta x/2 \right).
\end{equation}
Here $V_0$ is the optical potential depth, and $k = 2 \pi/\lambda$. Technically, the polarization is rotated by a combination of an electro-optical modulator (EOM) and fixed retardation plates. The two state-dependent standing wave potentials overlap if the linear polarization of both laser beams is parallel. This is the case for two different voltages applied to the EOM, i.e. at the beginning and at the end of a shifting operation. At the same time, the voltage range is limited to the range between these two values, so that the lattice can be separated by at most one potential well in a single shift. In order to perform the transport over several potential wells, the roles of \up\ and \down\ are exchanged in each step \cite{mandel_coherent_2003}.
The ramp changing the EOM voltage is approximately linear; the time of the ramp has been optimized for fast transport over one lattice site on the one hand and minimal excitation probability of the vibrational state \cite{mandel_coherent_2003} on the other hand. Using first-order perturbation theory approximation \cite{jaksch_entanglement_1999}, it can be shown that the excitation probability as a function of ramp time $\tau$ is proportional to $\mathrm{sinc}^2(\omega_{\mathrm{ax}}\tau/2)$. This approximation fits reasonably well to our measured data, from which an optimum ramp time of $19\,\mu$s corresponding to the second minimum of the measured $\mathrm{sinc}$-function has been deduced, limiting the probability to excite the vibrational states to less than 3\%.

\paragraph*{Coin operation}

After optical pumping to $\ket{0}=\ket{F=4, m_F=4}$, initialization and coin operations are performed by resonant microwave radiation between states \up\ and \down . The maximum Rabi frequency is $\Omega_0 = 2\pi \times 60\,$kHz. For all pulses, pulse area and phase can be independently controlled. This enables us to apply arbitrary single-particle rotations corresponding to any Pauli-spin operation $\hat{\sigma}_i$ with $i = x,y,z$. The first $\pi/2$-pulse ($\hat{\sigma}_x$-rotation with pulse area $\pi/2$) thereby applies the transformation 
\begin{eqnarray}
\up &\rightarrow& (\up + i\down )/\sqrt{2} \\
\down &\rightarrow& (\up - i\down )/\sqrt{2} . 
\end{eqnarray}
The following coin operations are Hadamard-type gates, realized as $3\pi/2$-pulses with phase $\pi/2$ with respect to the first pulse. The $3\pi/2$ pulse area can be decomposed into a $\pi/2$-pulse acting as the actual coin, and a $\pi$-pulse, acting as a spin echo for the internal states. The phase shift, compared to the first pulse, causes a $\hat{\sigma}_y$ rotation, i.e.~rotations around the imaginary axis of the Bloch sphere. In total this operation yields a Hadamard-type transform 
\begin{equation}
\hat{C}: \left \{ \begin{array}{c} \up \rightarrow (\up - \down)/\sqrt{2} \\ \down \rightarrow (\up + \down)/\sqrt{2} \end{array} \right. ,
\end{equation}
with spin echo included. The spin echo allows us to avoid the unwanted influence of dephasing due to temperature dependent differential light shifts, time-independent gradients of external magnetic or light fields without affecting the properties of the quantum walk. It extends the coherence time to roughly $0.8\,$ms.

During the sequence, apart from the first step, the roles of \up\ and \down\ are exchanged in each step. As a consequence the symmetry of the measured distribution is inverted compared to the case where the states are not exchanged, while the interference phenomenon of the quantum walk is not altered. Taking this into account yields the predictions we show in Fig.~\ref{fig:Fig1},\ref{fig:Fig2},\ref{fig:Fig3} and \ref{fig:Fig4}.

\paragraph*{Errors and imperfections in quantum walks}

The mechanisms limiting the performance of our quantum walk can be divided into two classes: First, external perturbations such as fluctuations of the ambient magnetic field, laser power, beam pointing, or spontaneous emission processes, lead to a limited coherence time even in the absence of a quantum walk sequence. These fluctuations cannot be canceled by the  spin-echo, limiting the coherence time without shift operation to 0.8\,ms.
Second, the operations involved in the quantum walk itself -- state preparation, coin and shift -- can be erroneous. For the state preparation and coin operation, imperfect microwave pulses with errors in amplitude or phase lead to an accumulation of unknown errors, and eventually the phase relation between populated sites is lost. For the shift operation, diabatic transfer of the atoms leads to excitation of higher vibrational states, which immediately results in a suppression of coherence, as two orthogonal modes of the wave function cannot interfere. We find an overall single-step fidelity of 96\% for a combined coin and shift operation.

The interplay of different decoherence sources together with systematic errors makes it difficult to model experimental imperfections and quantitatively extract the respective contributions from the measured data. The time reversal experiment, however, allows to illustratively distinguish the coherent fraction of atoms, being entirely refocused to the initial site of the lattice, from the incoherent rest, yielding a broad background. Using a simplified model, we have checked that decoherence effectively leads to a Gaussian background. A more detailed analysis, however, requires further theoretical investigation.

\paragraph*{Quantum state tomography in an optical lattice}
Our local quantum state tomography is a site-resolved extension of the tomography reported in \cite{rosenfeld_remote_2007}.
In our case, the population of state \down\ is measured by removing atoms in state \up\ from the trap by application of a laser beam resonant to the $\ket{F=4} \longrightarrow \ket{F^\prime = 5}$ transition of the Cs $D_2$-line. Subsequent imaging of the remaining atoms in the trap yields the population of the \down\ state in the lattice sites with an error $<5\%$. Applying proper microwave pulses just before the state-selective detection, the population of each eigenstate of the Pauli spin operators $\hat{\sigma}_i$, ($i=x,y,z$) can be measured. 

For two orthogonal eigenstates ($|e_{i,1}\rangle$, $|e_{i,2}\rangle$) belonging to the Pauli spin operator $\hat{\sigma}_{i}$ the sum of the corresponding population distributions follows 
\begin{equation}
P_{N,|e_{i,1}\rangle}(\xi)+P_{N,|e_{i,2}\rangle}(\xi)=P_{N}(\xi), 
\end{equation}
providing a consistency check for the measured data. This sum agrees with the overall, independently measured probability distribution within {$2.5$, $3.0$, $2.1$} times the statistical error for $i = x,y,z$, respectively. The deviations also reflect the sensitivity of the central lattice sites to small errors due to frequent matter wave interference.
Part of the deviation is due to technical limitations of our experimental setup, in particular concerning the long time stability as the measurement of each eigenstate population distribution takes approximately 30 minutes in our case.
 
In order to perform not only local but full quantum state tomography, also the off-diagonal elements of the position space density matrix of the final state have to be measured. This can, in principle, be done by applying $j$ shifting operations prior to the last coin operator which then allows to perform local measurements of the off-diagonal elements which have the form $\ket{x_i}\bra{x_i \pm j\,\lambda /4}$. However, this scheme is challenging in our case, as the sequence involves several shifting operations even for small quantum walk distributions, leading to significant decoherence. We stress, however, that the time-reversal of the quantum walk demonstrates the unitarity of the sequence as well as phase coherence of the matter wave interference across all populated lattice sites.

\bibliography{qwalk}

\end{document}